\documentstyle[aps,epsf,epsfig]{revtex}
\newcommand{\bq}{\begin{equation}}
\newcommand{\eq}{\end{equation}}
\newcommand{\bqn}{\begin{eqnarray}}
\newcommand{\eqn}{\end{eqnarray}}
\newcommand{\nb}{\nonumber}
\newcommand{\lb}{\label}

\begin{document}
\title{Gravitational Collapse of  Self-Similar Perfect Fluid
 in $2+1$ Gravity}
\author{A.Y. Miguelote \thanks{E-mail: yasuda@if.uff.br} $\; {}^a$ ,
N.A. Tomimura \thanks{E-mail: nazt@if.uff.br} $\;{}^a$ ,
 Anzhong Wang \thanks{ E-mail: aw@guava.physics.uiuc.edu} $\;{}^{b,\; c}$}
 \address{$^a$ Instituto de F\'{\i}sica, Universidade Federal Fluminense,
Av. Litor\^{a}nea s/n, Boa Viagem, 24210-340, Niter\'{o}i,
 RJ, Brazil\\
$^b$ Department of Physics, University of Illinois at
Urbana-Champaign, 
1110 West Green Street, Urbana, IL61801-3080, USA\\
$^c$ Departamento de F\' {\i}sica Te\' orica,
Universidade do Estado do Rio de Janeiro, 
Rua S\~ ao Francisco Xavier $524$, Maracan\~ a,
$20550-013$, Rio de Janeiro, RJ, Brazil}
\date{\today }
\maketitle

\begin{abstract}

Perfect fluid with kinematic self-similarity is studied in $2+1$
dimensional spacetimes with circular symmetry, and various exact
solutions to the Einstein field equations are given. In particular, 
these include all the solutions  of  dust and stiff perfect fluid 
with self-similarity of the first kind, and all the solutions of 
perfect fluid with a linear equation of state  and self-similarity 
of the zeroth  or second kind. It is found that some of these solutions 
represent gravitational collapse, and the final state of the collapse
can be either black holes or naked singularities.

\end{abstract}

\vspace{.6cm}


\section{Introduction}
 \lb{introd}
 \renewcommand{\theequation}{1.\arabic{equation}}
\setcounter{equation}{0}

One of the most outstanding problems in gravitation theory is the
final state of a collapsing massive star, after it has exhausted
its nuclear fuel. In spite of  numerous efforts over the last three
decades or so, because of the (mathematical) complexity of the problem
our understanding is still limited to several
conjectures, such as, the cosmic censorship conjecture
\cite{Penrose} and the hoop conjecture \cite{Thorne}. To the
former,   many counter-examples have been found \cite{Joshi},
although  it is still not clear whether  those particular examples
are  generic. To the latter, no counter-example has been found so
far in four-dimensional spacetimes, but it has been shown recently
that this is no longer true in five dimensions \cite{NM01}.

Lately,  Brandt {\em et al.} have studied gravitational collapse of perfect fluid
with kinematic self-similarities in four-dimensional spacetimes
\cite{Carlos02}, a subject that has been recently studied
intensively (for example, see  \cite{fluid} and references
therein.). In this paper, we shall investigate the same problem
but in 2+1 gravity \cite{Carlip}. The main motivation of such a
study comes from recent investigation of critical collapse of a
scalar field in 3D gravity \cite{PC00,Gar01,CF01,HW02}. It was found 
that in the 3D case the corresponding problem is considerably simplied
and can be studied analytically. In particular, Garfinkle first found a
class, say, $S[n]$, of exact solutions to Einstein-massless-scalar field
equations, and then Garfinkle and Gundlach studied their linear perturbations
and found that the solution with $n = 2$ has only one unstable mode \cite{Gar01}. 
By definition this is a critical solution, and the corresponding 
exponent, $\gamma$, of the black
hole mass,  
$$
M_{BH} \propto (p - p^{*})^{\gamma},
$$
is $\gamma = |k_{1}|^{-1} = 4/3$, where $k_{1}$ denotes the unstable mode.  
Although the exponent $\gamma$ is close to the one found numerically by 
Pretorius and Choptuik, which is $\gamma \sim  1.2 \pm 0.02$ (but not the 
one  of Husain and Olivier, $\gamma \sim  0.81$), this solution is 
quite different from the 
numerical one \cite{PC00}. Using different boundary conditions \footnote{
Comparing   the two sets of boundary conditions one 
will find that the only difference between them
is that, in addition to the ones imposed by Garfinkle 
and Gundlach \cite{Gar01}, Hirschmann, 
Wang and Wu \cite{HW02} further required that no matter 
field should come out of the already formed black holes. This
additional condition seems  physically quite reasonable, and has been widely used
in the studies of black hole perturbations \cite{Chandra83}.}, 
Hirschmann, Wang and Wu found that the solution with $n = 4$ has only one unstable
mode \cite{HW02}. As first noted by Garfinkle \cite{Gar01}, this $n = 4$ solution 
matches extremely well with the numerical critical
solution found by Pretorius and Choptuik \cite{PC00}. However, 
the corresponding exponent $\gamma$ now is given by
$\gamma = |k_{1}|^{-1} = 4$, which is significantly different from the numerical
ones.   In this paper we do not intend to solve these problems, but study some analytical 
solutions that represent gravitational collapse of perfect fluid in $2+1$ gravity.

It can be shown that the conception of {\em kinematic
self-similarities} given by Carter and Henriksen \cite{CH89} in
four-dimensional spacetimes can be easily  generalized to
D-dimensional spacetimes with the metric
 \bq
 \lb{1.1}
 ds^{2} = l^{2}\left\{\gamma_{ab}\left(x^{c}\right)dx^{a}dx^{b}
          -s^{2}\left(x^{c}\right)H_{ij}\left(x^{k}\right)
          dx^{i}dx^{j}\right\},
 \eq
 where $l$ is an unit constant with the dimension of length, so that all the
coordinates
 $x^{\mu}$ and metric coefficients $\gamma_{ab}$ and $H_{ij}$  are dimensionless.
 Here we use lowercase Latin indices, such as, $a, b, c, ..., g$, to run
 from $0$ to $1$, lowercase Latin indices, such as, $i, j, k, ...$, to run
 from $2$ to $D-1$, and Greek indices, such as, $\mu, \nu, ...$,
 to run from $0$ to $D-1$. Clearly, the above metric is invariant
 under the coordinate transformations
  \bq
  \lb{1.2}
  x^{a} = x^{a}\left({x'}^{c}\right),\;\;\;
  x^{i} = x^{i}\left({x'}^{j}\right).
  \eq
%Later we shall use this gauge freedom to fix the coordinates.

 On the other hand, the energy-momentum tensor (EMT) for a perfect fluid
 is given by
  \bq
  \lb{1.3}
  T_{\mu\nu} = \left(\rho + p\right)u_{\mu}u_{\nu} - pg_{\mu\nu},
  \eq
  where $u_{\mu}$ denotes the velocity of the fluid,  $\rho$ and
  $p$ are, respectively, its energy density and pressure. Using
  the coordinate transformations (\ref{1.2}) we shall choose the
  coordinates such that,
   \bq
   \lb{1.4}
   u_{\mu} =  \left(g_{00}\right)^{1/2}\delta^{0}_{\mu},\;\;\;\;
   g_{01} = 0.
   \eq
This implies that   the coordinates are comoving with the perfect fluid.
Then, metric (\ref{1.1}) can be cast in the form,
   \bq
   \lb{1.5}
   ds^{2} = l^{2}\left\{e^{2\Phi(t,r)}dt^{2} - e^{2\Psi(t,r)}dr^{2}
            - r^{2}S^{2}(t,r)H_{ij}\left(x^{k}\right)dx^{i}dx^{j}
            \right\}.
  \eq

Following Carter and Henriksen, we define kinematic
self-similarity by
 \bq
 \lb{1.6}
 {\cal{L}}_{\xi}h_{\mu\nu} = 2 h_{\mu\nu},\;\;\;\;
 {\cal{L}}_{\xi}u^{\mu} = -\alpha u^{\mu},
 \eq
where ${\cal{L}}_{\xi}$ denotes the Lie differentiation along the
vector field $\xi^{\mu}$, $\alpha$ is a {\em
dimensionless} constant,  and $h_{\mu\nu}$ is the project
operator, defined by
 $
 %\bq
 % \lb{1.7}
 h_{\mu\nu} \equiv g_{\mu\nu} - u_{\mu} u_{\nu}$.
 %\eq
When $\alpha = 0$, the corresponding solutions are said to have
self-similarity of {\em the zeroth kind}; when $\alpha = 1$, they
are said to have  self-similarity of {\em the first kind} (or {\em
homothetic self-similarity}); and when $\alpha \not= 0, 1$, they
are said to have self-similarity of {\em the second kind}.

Applying the above definition to metric (\ref{1.5}), we find
that the metric coefficients must take the form,
 \bq
 \lb{1.8}
 \Phi(t,r) = \Phi(x),\;\;\;
 \Psi(t,r) = \Psi(x),\;\;\;
 S(t,r) = S(x),
 \eq
where the self-similar variable $x$ and the vector field $\xi^{\mu}$ are given by
 \bqn
 \lb{1.9}
 & & \xi^{\mu}\frac{\partial}{\partial x^{\mu}}
   = \frac{\partial}{\partial t} + r\frac{\partial}
   {\partial r},\nb\\
 & & x = \ln(r) - t,\;\; (\alpha = 0),
 \eqn
for the zeroth kind, and
 \bqn
 \lb{1.10}
 & & \xi^{\mu}\frac{\partial}{\partial x^{\mu}}
   = \alpha t \frac{\partial}{\partial t} + r\frac{\partial}
   {\partial r},\nb\\
 & & x = \ln (r) - \frac{1}{\alpha}\ln(-t),\;\; (\alpha \not= 0),
 \eqn
 for the first ($\alpha = 1$) and second ($\alpha \not= 1$) kind,
 respectively.

As mentioned above, in this paper we shall study   perfect fluid
with kinematic self-similarity in $2 +1$-dimensional spacetimes
with circular symmetry, for which metric (\ref{1.5}) becomes
 \bq
 \lb{1.11}
   ds^{2} = l^{2}\left\{e^{2\Phi(x)}dt^{2} - e^{2\Psi(x)}dr^{2}
            - r^{2}S^{2}(x)d\theta^{2}\right\},
 \eq
where the hypersurfaces $\theta = 0, 2\pi$ are identified.
It should be noted that the   above metric is of invariance
under the following transformations,
 \bq
 \lb{rescaling}
 t = A \bar{t},\;\;\;
 r = B \bar{r},\;\;\; 
 g_{\mu\nu} = C^{2}\bar{g}_{\mu\nu},
 \eq
 for  self-similar solutions of the first and second kinds, and
 \bq
 \lb{rescalingb}
 t = \bar{t} + A,\;\;\;
 r = B \bar{r},\;\;\; 
 g_{\mu\nu} = C^{2}\bar{g}_{\mu\nu},
 \eq 
 for self-similar solutions of the zeroth kind, 
where $A,\; B$ and $C$  are arbitrary constants.

On the other hand, to have circular symmetry, some physical 
and geometrical conditions needed to be
imposed \cite{Barnes}. In general this is not trivial. As a matter of fact,  
only when the symmetry axis is free of spacetime singularity, do we know how to 
impose these conditions. Since in this paper we are mainly interested in gravitational
collapse, we shall assume that the axis is regular at the beginning of the
collapse, so that we are sure that the singularity to be formed later on the axis
is due to the collapse of the fluid. Following \cite{Fatima} we impose 
the following conditions:

(i) There must exist a symmetry axis, which can be expressed as  
\bq
\lb{cd1}
X \equiv \left|\xi^{\mu}_{(\theta)}\xi^{\nu}_{(\theta)}g_{\mu\nu} 
\right| = \left|g_{\theta\theta}\right| \rightarrow 0,
\eq
as $r \rightarrow 0^{+}$, where we have chosen the radial coordinate such 
that the axis  is located at $r = 0$, and $\xi^{\mu}_{(\theta)}$ is the Killing vector
with a close orbit, given by $\xi^{\alpha}_{(\theta)}\partial_{\alpha} = 
\partial_{\theta}$.

(ii) The spacetime near the symmetry axis is locally flat, which can be
written as \cite{Kramer80}
\bq
\lb{cd2}
\frac{X_{,\alpha}X_{,\beta} g^{\alpha\beta}}{4X} \rightarrow
- 1,
\eq
as  $r \rightarrow 0^{+}$, where $(\;)_{,\alpha} \equiv \partial (\;)/\partial 
x^{\alpha}$. Note that solutions failing to satisfy this condition are sometimes
acceptable. For example, when the left-hand side of the above equation 
approaches a finite constant, the singularity at $r = 0$ can be related to
a point-like particle  \cite{VS}. However, since here we are mainly interested 
in gravitational collapse, in this paper we shall assume that this condition holds 
strictly at the beginning of the collapse.

(iii) No closed timelike curves (CTC's). In spacetimes with circular symmetry,
CTC's can be easily introduced. To ensure their absence we assume that
\bq
\lb{cd3}
g_{\theta\theta} = \xi^{\mu}_{(\theta)}\xi^{\nu}_{(\theta)}g_{\mu\nu} < 0,
\eq
holds in the whole spacetime. 

In addition to these conditions, it is usually 
also required that the spacetime be asymptotically flat in the radial direction. 
However, since we consider solutions with self-similarity, this condition
cannot be satisfied by such solutions, unless we restrict the validity of
them only up to a maximal radius, say, $r = r_{0}(t)$. 
Then, we need to join the solutions
with others in the region $r > r_{0}(t)$, which are 
asymptotically flat as $r \rightarrow \infty$. 
In this paper, we shall not consider such a possibility, and simply assume
that  the self-similar solutions are valid in the whole spacetime.

It should be noted that   the regularity
conditions (\ref{cd1})-(\ref{cd3}) are invariant 
under the transformations of Eq.(\ref{rescaling})
or Eq.(\ref{rescalingb}).  
Using these transformations, we shall further assume that
\bq
\lb{cd4}
\Phi(t, 0) = 0,
\eq
that is, the timelike coordinate $t$ measures the proper time on the axis.

Moreover, in the analytical studies of critical collapse, one is usually first to
find some particular solutions by imposing certain symmetries, such as,
homothetic  self-similarity, and then study their perturbations,  
because by definition critical solutions have one and only one unstable 
mode. In this paper we shall be mainly concerned with the first question, and
leave the study of linear perturbations to another occasion.
In particular, the rest of the   paper is organized as follows: Exact solutions
of the Einstein field equations with self-similarity of the first,
second, and zeroth kinds will be given and studied, respectively, in Secs.
$II$, $III$, and $IV$, while in Sec. $V$, our main conclusions are
presented. There are also two appendices $A$ and $B$, where in
Appendix $A$ the general expression of the Einstein tensor, among
other things, for spacetimes with self-similarities of the zeroth,
first and second kinds is given, while in Appendix $B$ the
apparent horizon is defined in terms of the self-similar
variables.

\section{Self-Similar Solutions of the First Kind}
\lb{firstkind}

\renewcommand{\theequation}{2.\arabic{equation}}
\setcounter{equation}{0}

In this section, we shall study solutions with self-similarity of
the first kind. Substituting Eqs.(\ref{1.3}), (\ref{1.4}) and
(\ref{A.12}) with $\alpha = 1$ into the Einstein field equations $
G_{\mu\nu} =  \kappa T_{\mu\nu}$, where $\kappa $ is the Einstein
coupling constant, we find that
 \bqn
 \lb{2.2a}
   & & y_{,x} - (1+y)\left(\Psi_{,x} - y\right) - y\Phi_{,x} = 0,\\
   \lb{2.2b}
   & & \Phi_{,xx}  + \Phi _{,x}\left( \Phi _{,x}-\Psi _{,x}-y-2\right)
   \nonumber \\
   & & \;\;\;\;\;\; -e^{2\left( x+\Psi -\Phi \right) }
   \left[\Psi_{,xx}-\Psi _{,x}\left( \Phi _{,x}
   -\Psi _{,x}+y\right) \right] = 0,
  \eqn
 and
  \bqn
  \lb{2.3}
  \rho &=&\frac{ye^{-2\Psi }}{  l^{2} r^2}\left(\Psi _{,x}
  e^{2(x+\Psi -\Phi)}-\Phi _{,x}\right),  \nb\\
  p &=& - \frac{\left( y+1\right) e^{-2\Psi }}
 {  l^{2} r^2}\left(\Psi _{,x}
  e^{2(x+\Psi -\Phi)}-\Phi _{,x}\right),
 \eqn
 where $y \equiv S_{,x}/S$.  Note that in writing the above expressions
 we had set $\kappa = 1$. In the rest of this paper we shall continuously 
 choose units so that this is true.   Clearly, to determinate
 the metric coefficients completely, we need to impose the equation of state
 for the perfect fluid. In general, it takes the form \cite{ST83},
 \bq
 \lb{2.3a}
 \rho = \rho(T,  \Sigma), \;\;\;\;
 p = p(T,  \Sigma),
 \eq
 where $T$ is the temperature of the system and $\Sigma$ the entropy.
However, in some cases  the system is 
weakly dependent on $T$, so the  equation of
 state can be written approximately as $p = p(\rho)$. In the following we
 shall show that in the latter case the only equation of
 state that is consistent with the symmetry of homothetic
 self-similarity is the one given by
  \bq
 \lb{2.4}
 p =k \rho,
 \eq
where $k$ is an arbitrary constant. To show   this, let us
first write Eq.(\ref{2.3}) in the form
    \bqn
   \lb{2.3aa}
    \rho &=& \frac{f(x)}{r^{2}},\\
   \lb{2.3ab}
    p &=& \frac{g(x)}{r^{2}}.
    \eqn
Then, from Eq. (\ref{2.3aa}) we find that
   \bq
   \lb{2.3ac}
   x = x\left(r^{2}\rho\right),\;\;\;\;
   \frac{dx}{d\left(r^{2}\rho\right)} =  \frac{1}{f'(x)},
   \eq
where a prime denotes the ordinary differentiation with respect to
$x$. Inserting Eq.(\ref{2.3ac}) into Eq.(\ref{2.3ab}) we find that
   \bq
   \lb{2.3ad}
   p = \frac{g\left(x\left(r^{2}\rho\right)\right)}{r^{2}},
   \eq
which shows that in general  $p$ is a function of $r$ and $\rho$.
Taking partial derivative of the above equation with respect to
$r$, and then setting it to zero, we find that
   \bqn
   \lb{2.3ae}
   \frac{\partial p(r, \rho)}{\partial r} &=& - 2\frac{g(x)}{r^{3}} +
   \frac{1}{r^{2}}\frac{dg(x)}{dx}\frac{dx}{d\left(r^{2}\rho\right)}
   \frac{\partial\left(r^{2}\rho\right)}{\partial r}\nb\\
   &=& \frac{2fg}{r^{3}f'}\left(\frac{g'}{g} - \frac{f'}{f}\right) =
   0,
   \eqn
which gives $g(x) = k f(x)$. Then,   Eq.(\ref{2.4}) follows.
In the following we shall further assume $ 0 \le k \le 1$ so
that all the energy conditions hold \cite{HE73} and that the
pressure is non-negative.

The combination of Eqs.(\ref{2.3}) and (\ref{2.4}) immediately
yields
 \bq
 \lb{2.5}
 y = - \frac{1}{1 + k}.
 \eq
Then,   Eqs.(\ref{2.2a}) and (\ref{2.5}) have the
solutions,
  \bqn
  \lb{2.6a}
  & & \Phi = k\left(\frac{x}{1 + k} + \Psi\right) + \Phi_{0},\nb\\
  \lb{2.6b}
  & & S(x) = S_{0} e^{- x/( 1 +k)},
  \eqn
while Eq.(\ref{2.2b}) becomes
  \bqn
  \lb{2.6c}
  & & k \Psi _{,xx}+k \left( \Psi _{,x}+\frac 1{k +1}\right)
  \left[ \left( k -1\right) \Psi _{,x}-1\right]  \nonumber \\
  &&-e^{2\left[ \frac x{k +1}+\left( 1-k \right) \Psi
  - \Phi_{0}\right]}\left[ \Psi _{,xx}+\left( 1-k \right) \Psi _{,x}
  \left( \Psi _{,x}+ \frac 1{k +1}\right) \right] =0,
  \eqn
 where $\Phi_{0}$ and $S_{0}$ are  integration constants.

 \subsection{ Stiff Fluid ($k = 1$)}

 When $k = 1$, i.e., the stiff fluid, Eq.(\ref{2.6c}) has the general
 solution,
 \bqn
 \lb{2.7a}
 \Psi(x) &=& q\ln\left(1 - e^{x - x_{0}}\right) 
   - \frac{1}{2}\left(x - x_{0}\right) + \Psi_{0},\nb\\
 \Phi(x) &=& q\ln\left(1 - e^{x - x_{0}}\right) 
   + x_{0} + \Psi_{0},\nb\\
S(x) &=& S_{0}e^{-x/2},   
 \eqn
where $\Psi_{0}, x_{0}, S_{0}$ and $q$ are the integration
constants, and $x_{0} \equiv 2\Phi_{0}$. Imposing the regularity 
conditions of Eqs.(\ref{cd1})-(\ref{cd3}) and the gauge condition
(\ref{cd4}), we find
that 
\bq
\lb{2.7aaa}
S_{0} = 2 e^{-x_{0}/2},\;\;\;\;
\Psi_{0} = - x_{0},
\eq
while $x_{0}$ and $q$ are arbitrary and cannot be removed by the transformations of
Eq.(\ref{rescaling}).  Then, the general solution  is given
by
 \bq
 \lb{2.7b}
 ds^{2} = l^{2}\left\{ \left(1 - e^{x-x_{0}}\right)^{2q}\left[dt^{2}
 - e^{-(x + x_{0})}dr^{2}\right] - 4r^{2}e^{-(x + x_{0})}d\theta^{2}\right\},
 \eq
and the corresponding energy density of the fluid reads,
 \bq
 \lb{2.7c}
 \rho =  p = \frac{1-2q}{4  l^{2}(-t)^{2}
 \left(1 - e^{x-x_{0}}\right)^{2q}}.
 \eq
It should be noted that in writing Eq.(\ref{2.7b}), we had
implicitly assumed that $x - x_{0} \le 0$. On the other hand, from
Eq.(\ref{2.7c}) we can see that we must have $q < 1/2$ in order to
have $\rho \ge 0$, a condition that we shall assume in the rest of
the paper. The metric is singular on the hypersurface $x = x_{0}$, 
and depending on the values of $q$, the nature of the singularity  
is different.
 
 \subsubsection{$0 < q < 1/2$}

In this case the singularity is a curvature one and
marginally naked. In fact, from Eq.(\ref{B.9}) we find that
 \bqn
 \lb{2.7da}
 \theta_{l} &=& \frac{fe^{x}}{2l^{2}r\left(1 -
 e^{x-x_{0}}\right)^{q}}\left[e^{(x_{0} - x)/2} - 1\right],\nb\\
 \theta_{n} &=& - \frac{ge^{x}}{2l^{2}r\left(1 -
 e^{x-x_{0}}\right)^{q}}\left[e^{(x_{0} - x)/2} + 1\right],\nb\\
 \eqn
from which  we can see that $\theta_{n} < 0$ for any given $x \in
[0, x_{0}]$, and $\theta_{l}$ is positive  for $x < x_{0}$, zero only on 
the surface $x = x_{0}$. Thus, now the hypersurface $x = x_{0}$ is a marginally
trapped surface and the  spacetime singularity located there
is marginally naked. On the other hand, for
any given hypersurface $x = Const$., say, $C$, its normal vector is given by
\bq
\lb{2.7d}
n_{\alpha} \equiv \frac{\partial(x - C)}{\partial x^{\alpha}}
= \frac{1}{r}\left(e^{x}\delta^{t}_{\alpha} + \delta^{r}_{\alpha}\right),
\eq
from which we find
\bq
\lb{2.7daaa}
n_{\alpha}n_{\beta}g^{\alpha\beta} = - \frac{e^{x+x_{0}}}{l^{2}r^{2}}
\left(1 - e^{x - x_{0}}\right)^{1-2q}.
\eq
Clearly, on the hypersurface $x = x_{0}$ the normal vector, $n_{\alpha}$,
becomes null. That is, the nature of the singularity is null.
The corresponding Penrose diagram is given by Fig. 1.

%%%%%%%%%%%%%%%%%%%%%%%%%%%%%%%%%%%%%%%%%%%%%%%%%%%%%%%%%%%%%%%%%%%%%%%%%%%%
%%
 \begin{figure}[htbp]
 \begin{center}
 \label{fig1}
 \leavevmode
  \epsfig{file=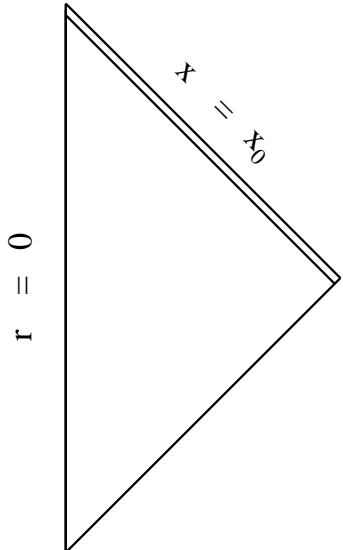,width=0.25\textwidth,angle=0}
 \caption{The Penrose diagram for the solutions given  by
 Eq.(\ref{2.7b}) with $0 < q < 1/2$.  The  spacetime is singular
 on the double line $x = x_{0}$, which is a null surface and on which
 the expansion of the outgoing radial null geodesics is zero,
 $\left.\theta_{l}(t,r)\right|_{x = x_{0}} = 0$, but $\theta_{n}(t,r) < 0$
 in the whole spacetime, including the hypersurface $x = x_{0}$.
  }
 \end{center}
 \end{figure}
%%%%%%%%%%%%%%%%%%%%%%%%%%%%%%%%%%%%%%%%%%%%%%%%%%%%%%%%%%%%%%%%%%%%%%%%%%%%

 \subsubsection{$q = 0$}

%{\bf Case B) $\;   q = 0 $}: 

In this case the metric is free of spacetime and coordinate 
singularities on the hypersurface 
$x = x_{0}$, and  valid in the whole region $t \le 0,\; r \ge 0$. However,
now the spacetime becomes  singular on the hypersurface $t = 0$ 
[cf. Eq.(\ref{2.7c})]. From Eq.(\ref{2.7da}), which is also valid for $q = 0$,
 we can see that this singularity 
is not naked, and always covered by the apparent horizon formed on the
hypersurface $x = x_{0}$. In the region $x > x_{0}$, which is denoted as 
Region $I$ in Fig. 2,  Eq.(\ref{2.7da}) shows
that $\theta_{l}$ becomes negative and $\theta_{l} \theta_{n} > 0$, that is, 
all the rings of constant $t$ and $r$ now
become trapped. Thus, in this case  Region $I$ can be considered as the interior
of a black hole, which is formed by the gravitational collapse of the stiff
fluid in Region $II$. The corresponding Penrose diagram is given by Fig. 2.

%%%%%%%%%%%%%%%%%%%%%%%%%%%%%%%%%%%%%%%%%%%%%%%%%%%%%%%%%%%%%%%%%%%%%%%%%%%%
%%
 \begin{figure}[htbp]
 \begin{center}
 \label{fig2}
 \leavevmode
  \epsfig{file=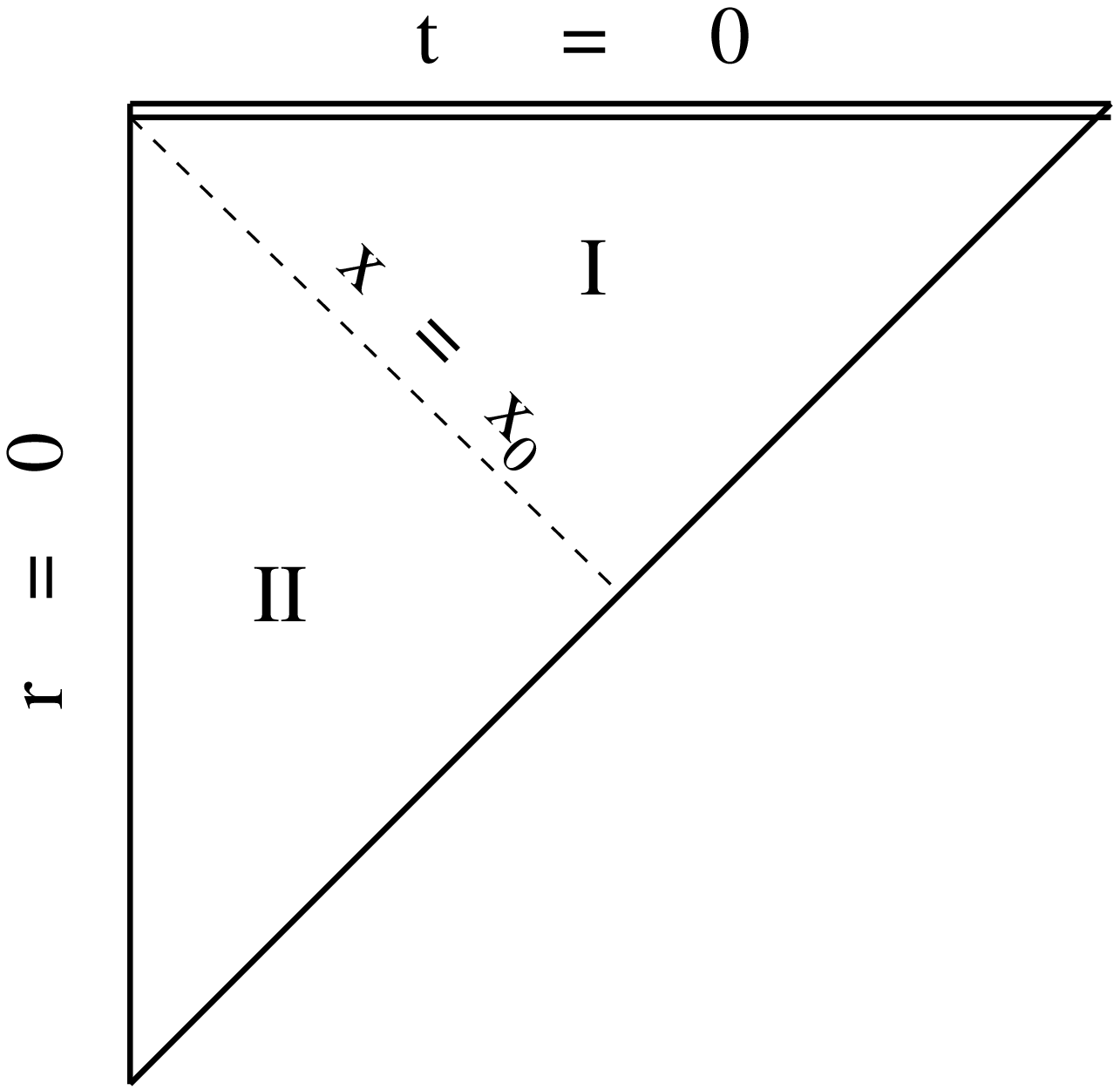,width=0.35\textwidth,angle=0}
 \caption{The Penrose diagram for the solutions given  by
 Eq.(\ref{2.7b}) with $q = 0$.  The  spacetime is singular
 on the double line $t = 0$. All the rings of constant $t$ and $r$ are
 trapped in Region $I$ but not in Region $II$, where
 $I = \left\{x^{\alpha}: x > x_{0}, t < 0, r \ge 0\right\}$ and
 $II = \left\{x^{\alpha}: x < x_{0}, t < 0, r \ge 0\right\}$.
 The hypersurface $x = x_{0}$ is a null surface and represents 
 an apparent horizon.}
 \end{center}
 \end{figure}
%%%%%%%%%%%%%%%%%%%%%%%%%%%%%%%%%%%%%%%%%%%%%%%%%%%%%%%%%%%%%%%%%%%%%%%%%%%%

It should be noted that apparent horizons and black holes are usually defined
in asymptotically flat spacetimes \cite{HE73}. As mentioned previously, the
spacetimes considered here are not asymptotically flat, therefore, strictly 
speaking the above definitions are, in some sense, the generalization of those
given in \cite{HE73}. As a matter of fact,  to be distinguishable with 
the asymptotical case, Hayward called such apparent horizons as trapping 
horizons and defined black holes by future outer trapping horizons
\cite{Hay94}. For the sake of simplicity and without causing any confusions, 
following \cite{Gar01,CF01,HW02} we shall continuously use the notions, 
apparent horizons.

\subsubsection{$q < 0$}

In this case the spacetime is free of
curvature singularity on the hypersurface $x = x_{0}$, although we
do have a coordinate one there. Then, to have a geodesically
maximal spacetime, we need to extend it beyond this surface.
Introducing two null coordinates $u$ and $v$ via the relations
 \bqn
 \lb{2.7e}
 t &=& - \frac{1}{2}\left[\left(-u\right)^{n} +
 \left(-v\right)^{n}\right]^{2},\nb\\
 r &=& \frac{1}{2}e^{x_{0}}\left[\left(-u\right)^{n} -
 \left(-v\right)^{n}\right]^{2},
 \eqn
where
 \bq
 \lb{2.7f}
 n \equiv \frac{1}{2q + 1},
 \eq
we find that the metric (\ref{2.7b}) becomes
 \bq
 \lb{2.7g}
 ds^{2} = l^{2}\left\{2 e^{2\sigma(u,v)}dudv 
 - R^{2}(u,v)d\theta^{2}\right\},
 \eq
with
 \bqn
 \lb{2.7h}
 \sigma(u,v) &=& \left(1-2q\right)\ln\left[\left(-u\right)^{n} +
                \left(-v\right)^{n} \right] + \sigma_{0},\nb\\
R(u,v)&=&  \left(-u\right)^{2n} - \left(-v\right)^{2n},\nb\\
  \sigma_{0} &\equiv& \frac{1}{2}\ln\left(2n^{2}4^{2q}\right),
\eqn 
and the corresponding velocity and energy density of the stiff fluid are 
given, respectively, by
 \bqn
 \lb{2.7ha}
 u_{\mu} &=& \frac{l e^{\sigma}}{\sqrt{2}}\left(u_{0}\delta^{u}_{\mu} 
 + \frac{1}{u_{0}}\delta^{v}_{\mu}\right),\;\;\;\;
 u_{0} \equiv \left(\frac{u}{v}\right)^{(n-1)/2},\nb\\
 \rho &=&  \frac{\rho_{0}(uv)^{n-1}}{\left[(-u)^{n} +
 (-v)^{n}\right]^{ 6 - 2/n}},\;\;\;
 \rho_{0} = \frac{2n(2n - 1)}{  l^{2}}e^{-2\sigma_{0}}.
 \eqn
From Eq.(\ref{2.7e})
we can see that the region $t\le 0, \; r \ge 0, \; x \le x_{0}$ in
the ($t, r$)-plane has been mapped into the region $u \le 0, \; v \le 0$ and
$v \ge u$ in the ($u,v$)-plane, which will be referred as to region
$II$, as shown in Fig. 3. The null hypersurface $x = x_{0}$, as can be seen
from Eq.(\ref{2.7daaa}), is mapped to the one $v = 0$.
 Region $I$, where $u \le 0,\; v \ge
0,\; |u| \ge v$, represents an extended region. In this extended
region, the metric is real only for $ {n} \not= (2m +1)/(2j)$,
where $j$ and $m$ are integers.  When $ {n} = (2m +1)/(2j)$, a
possible extension may be given by replacing $-v$ by $|v|$.
However, this extension is not analytical. In fact, no analytical extension
exists in this case. The only case where the extension is analytical is the
one where $n$ is an integer. When the extension is not analytical, it is
also not unique. Thus, to have an unique extension in the following we shall
assume that $n$ is an integer.  
On the other hand, from Eq.(\ref{B.5}) we find that
\bq
\lb{2.7hb}
\theta_{l} = \frac{2ne^{-2\sigma}}{l^{2}R}(-v)^{2n-1},\;\;\;
\theta_{n} = -\frac{2ne^{-2\sigma}}{l^{2}R}(-u)^{2n-1},
\eq
where $R \equiv r S$. From the above expressions  
we can see that $\theta_{n}$ is always negative in both regions
$I$ and $II$,   while $\theta_{l}$ is positive in Region $II$,  
zero on the hypersurface $v = 0$, and negative in the extended region, $I$.  
Near the surface $v = 0$ the only non-vanishing component of the energy-momentum
tensor is given by
\bq
\lb{2.7h3}
T_{uu} = \frac{l^{2}\rho_{0}e^{2\sigma_{0}}}{(-u)^{2}},
\eq
which represents the energy flow moving from Region $II$ to Region $I$ along the
hypersurfaces $u = Const$. To study the above solutions further, let us consider the
two cases $n = 2m + 1$ and $n = 2m$, separately, where  $m = 1, 2, 3, ...$.
 
When $n$ is an odd integer, i.e.,  $n = 2m + 1$, from Eq.(\ref{2.7ha}) we
can see that the spacetime is singular on the hypersurface $u =
-v$  in the extended region $I$, on which we have $R(u,v) = 0$. On
the other hand, all the rings of constant $t$ and $r$   are trapped
in this region, as we can see from Eq.(\ref{2.7hb}), now we  have
$\theta_{l}(v > 0) < 0$ and $\theta_{l}\theta_{n} > 0$.
 Thus, region $I$ can be considered as the interior
of the black hole that is formed from the gravitational collapse
of the fluid in region $II$. The corresponding Penrose diagram is 
that of Fig. 3.

%%%%%%%%%%%%%%%%%%%%%%%%%%%%%%%%%%%%%%%%%%%%%%%%%%%%%%%%%%%%%%%%%%%%%%%%%%%%
%%
 \begin{figure}[htbp]
 \begin{center}
 \label{fig3}
 \leavevmode
  \epsfig{file=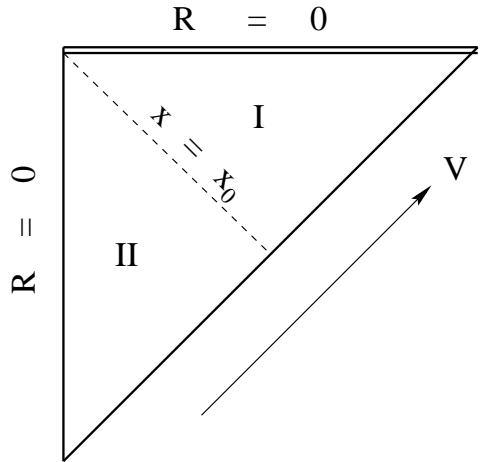,width=0.35\textwidth,angle=0}
 \caption{The Penrose diagram for the solutions given  by
 Eqs. (\ref{2.7g}) and (\ref{2.7h}) with $n$ being an integer. 
 The rings of constant $t$ and $r$ 
  are closed trapped surfaces in Region $I$, but not in region $II$. The dashed
  line $v = 0$, or $x = x_{0}$, 
  represents an apparent horizon.
  When $n$ is an odd integer, the spacetime is singular on the horizontal double 
  line $R = 0$, while when $n$ is an even integer, the spacetime has an angular defect
  there.}
 \end{center}
 \end{figure}
%%%%%%%%%%%%%%%%%%%%%%%%%%%%%%%%%%%%%%%%%%%%%%%%%%%%%%%%%%%%%%%%%%%%%%%%%%%%
%%

When $n$ is an even integer, say, $n = 2m$, from Eq.(\ref{2.7ha}) we
can see  that no spacetime curvature singularity is developed on the
axis $R = 0$ in the extended region $I$, although all the  rings of 
constant $t$ and $r$ are also trapped in this region, as now we still
have $\theta_{l}(v > 0) < 0$ and $\theta_{l}(v > 0)\theta_{n}(v > 0) > 0$.
However, the local flatness condition (\ref{cd2}) is not satisfied there. In
fact, it can be shown that now we have
\bq
\lb{2.7hc}
\frac{X_{,\alpha}X_{,\beta}g^{\alpha\beta}}{4X} \rightarrow + 1,
\eq
as $v \rightarrow - u$. Thus, unlike that on the axis $R = 0$ in Region $II$, 
where the local flatness condition is  satisfied, now the spacetime on 
the axis $R = 0$ in region $I$ has angle defect.
The corresponding Penrose diagram is 
also given by Fig. 3, but now the horizontal double line  $R = 0$,
instead of representing a  curvature singularity, now represents  
an angle-defect-like singularity \cite{VS}. 

Moreover, in the extended region $I$, where $uv < 0$, the function $\rho$ 
becomes negative, and the three-velocity $u_{\mu}$ becomes imaginary, 
as we can see from Eq.(\ref{2.7ha}). A close investigation shows that 
the energy-momentum tensor in this extended region  actually takes 
the form
 \bq
 \lb{2.7ja}
 T_{\mu\nu} = \bar{\rho}\left(2 r_{\mu}r_{\nu} + g_{\mu\nu}\right),
 \eq
where
\bqn
\lb{2.7jb}
r_{\mu} &=& \frac{le^{\sigma}}{\sqrt{2}}\left(r_{0}\delta^{u}_{\mu} 
 - \frac{1}{r_{0}}\delta^{v}_{\mu}\right),\;\;\;
 r_{0} \equiv \left|\frac{u}{v}\right|^{(2m-1)/2},\nb\\
 \bar{\rho} &=& \frac{\rho_{0}\left|uv\right|^{2m-1}}{\left(u^{2m} +
 v^{2m}\right)^{(6m-1)/m}},\; (n = 2m),
 \eqn
from which we find that
\bq
\lb{2.7jc}
g^{\mu\nu} r_{\mu}r_{\nu} = -1.
\eq
Thus, in the extended region the source is no more a stiff fluid. 
Introducing the unit vectors,
\bq
\lb{2.7jd}
u_{\mu} = \frac{le^{\sigma}}{\sqrt{2}}\left(r_{0}\delta^{u}_{\mu} 
 + \frac{1}{r_{0}}\delta^{v}_{\mu}\right),\;\;\;
 \theta_{\mu} = l R\delta^{\theta}_{\mu},
 \eq
where $u_{\nu} u^{\nu} = 1$ and $\theta_{\nu}\theta^{\nu} = -1$, 
we find that Eq.(\ref{2.7ja}) reads
 \bqn
 \lb{2.7je}
 T_{\mu\nu} &=& {\rho}u_{\mu}u_{\nu} - p_{r}r_{\mu}r_{\nu}  
 - p_{\theta}\theta_{\mu}\theta_{\nu},\nb\\
 \rho &=& - p_{r} = p_{\theta} = 
 \frac{\rho_{0}\left|uv\right|^{2m-1}}{\left(u^{2m} +
 v^{2m}\right)^{(6m-1)/m}}.
 \eqn
Thus, the source in region $I$ now becomes an anisotropic fluid
with its energy density $\rho$ and two principal pressures, $p_{r}$ and 
$p_{\theta}$, in the direction of , respectively, $r_{\mu}$ and $\theta_{\mu}$.
Note that although the pressure in the $r_{\mu}$ 
direction is negative, the anisotropic fluid satisfies all the three
energy conditions \cite{HE73}.   With this odd
feature, it is not clear whether the corresponding solution can be
interpreted as representing  gravitational collapse of the stiff fluid.
In fact, if we consider the fluid from its microscopic point of view, we
might be able to rule out such a change in the equation of state across
the hypersurface $v = 0$ \cite{FMS}, a problem that 
is under our current investigation.

Before turning to study other cases, we would like to note that
it is well-known that a still perfect fluid is energetically equivalent to
a massless scalar field when the velocity of the fluid is irrotational
$\nabla_{[\mu}u_{\nu]} = 0$ and the massless scalar field is timelike
$\phi_{,\alpha} \phi_{,\beta}g^{\alpha\beta} > 0$. Clearly, these
conditions are fulfilled in Region $II$. In fact, comparing
the solutions given by Eq.(\ref{2.7h}) with the corresponding massless scalar
field ones found in \cite{HW02}, one can find that 
they are actually the same in Region $II$. However, as shown lately in 4-dimensional case  
\cite{Cho02}, the spacetime across the horizon $v = 0$ can be quite
different. Our analysis given above and the one given in \cite{Gar01,HW02}
show that this is also the case in $2+1$ Gravity. In \cite{Cho02}, 
it was also shown that the linear perturbations in these two cases are
different.  Thus, it would be very interesting to study the linear perturbations
of the above solutions in terms of stiff fluid.  

\subsection{Dust Fluid  $(k = 0)$}

When $k = 0$, the pressure of the perfect fluid vanishes,
i.e., the dust fluid. Then, it can  be shown that in this case the
general solution is given by
 \bqn
 \lb{2.15}
 \Phi(x) &=& \Phi_{0},\nb\\
 \Psi(x) &=& \ln\left(1 - e^{x-x_{0}}\right)
 - x  + \Psi_{0},\nb\\
 S(x) &=& S_{0} e^{-x},
 \eqn
for which we find
 \bq
 \lb{2.16}
 \rho = \frac{e^{-2\Phi_{0}}}{  l^{2} (-t)^{2}
        \left(1 -e^{x-x_{0}}\right)},\;\;\;
 p = 0.
 \eq
Since now we have $R(t, r) = rS(x) = -S_{0} t$, we can see that these
solutions are  Kantowski-Sachs like  \cite{Kramer80}, and may be
interpreted as representing cosmological models. Because in this paper 
we are mainly interested in gravitational collapse, in the following we shall 
not consider this case in any more details.

\subsection{Perfect Fluid With $k \not= 0, 1$}

 When $k \not=0, 1$, introducing the function $Z(x)$ by
 \bq
 \lb{2.7}
 Z(x) = {\rm exp}\left\{2\left[\frac{x}{1 + k}
 + \left(1-k\right)\Psi - \Phi_{0}\right]\right\},
 \eq
 we find that Eq.(\ref{2.6c}) can be written in the form,
 \bqn
 \lb{2.8}
 2Z\left(k -Z\right)Z_{,xx}
 &+& \left(Z-3k\right){Z_{,x}}^2
 +2Z^{2}Z_{,x} \nb\\
 &+& \frac{4k}{\left(k +1\right)^2} Z^2
 \left(k ^2  - Z\right) = 0.
 \eqn
We are not able to find the general solution of this equation,
but a particular one  given by
 \bqn
 \lb{2.10}
 \Phi(x) &=& \Phi_{0}, \nb\\
 \Psi(x) &=&-\frac x{k +1} + \frac{\Phi_{0}}{1 - k}, \nb\\
 S(x) &=& S_{0}e^{-\frac x{k +1}}.
 \eqn
Then, it can be shown that the conditions of Eqs.(\ref{cd1})-(\ref{cd4})
are fulfilled, provided that
\bq
\lb{2.10a}
\Phi_{0} = 0,\;\;\;\;
S_{0} = \frac{1+k}{k},\;\;\; k > 0.
\eq
Thus,  the corresponding metric takes the form
 \bq
 \lb{2.11}
 ds^{2} = l^{2}\left\{dt^{2} - e^{-2x/(1 +k)}
 \left[dr^{2} + \left(\frac{1+k }{k}\right)^{2}
 r^{2}d\theta^{2}\right]\right\},
 \eq
and the pressure and energy density of the fluid are given by
 \bq
 \lb{2.12}
 p = k\rho = \frac{k}{  l^{2}(1 + k)^{2} (-t)^{2}}.
 \eq
From the above expression we can see that   the spacetime
is always singular at $t = 0$. However, this singularity is not naked. 
In fact, from Eq.(\ref{B.9}) we
find that  
\bqn
\lb{2.12a}
\theta_{l} &=& \frac{fe^{x/(1+k)}}{l^{2}(1+k)r}
\left(k - e^{k x/(1+ k)}\right),\nb\\
\theta_{n} &=& - \frac{ge^{x/(1+k)}}{l^{2}(1+k)r}
\left(k + e^{k x/(1+ k)}\right),
\eqn
from which we can see that $\theta_{n}$ is always negative, and
\bq
\lb{2.12b}
\theta_{l}(t,r) = \cases{ > 0, \;\;\;  x < x_{0},\cr
 = 0,  \;\;\; x = x_{0},\cr
 < 0, \;\;\; x > x_{0},\cr}
 \eq
 where $x_{0}$ is given by
 \bq
 \lb{2.12c}
 x_{0} \equiv \frac{1+k}{k}\ln(k).
 \eq
Thus, the singularity  at $t = 0$ is always covered by the 
apparent horizon localized on the hypersurface $x = x_{0}$.
In the region where $t < 0$ and $x > x_{0}$, the rings of constant
$t$ and $r$ are all trapped, as now we have $\theta_{l} < 0$
and $\theta_{l} \theta_{n} > 0$. Therefore, in the present
case the collapse of the fluid always forms black holes.
Moreover, the normal vector
$n_{\alpha}$ to the hypersurface $x = x_{0}$ is still given 
by Eq.(\ref{2.7d}) but now with 
\bq
\lb{2.12d}
n_{\alpha}n_{\beta}g^{\alpha\beta} = - \frac{e^{2x_{0}/(1+k)}}
{l^{2} r^{2}}\left(1 - k^{2}\right)
< 0,
\eq
since now we have $0 < k < 1$. That is, the apparent horizon 
is always timelike and the corresponding Penrose diagram 
 is given by Fig. 4.

%%%%%%%%%%%%%%%%%%%%%%%%%%%%%%%%%%%%%%%%%%%%%%%%%%%%%%%%%%%%%%%%%%%%%%%%%%%%
%%
 \begin{figure}[htbp]
 \begin{center}
 \label{fig4}
 \leavevmode
  \epsfig{file=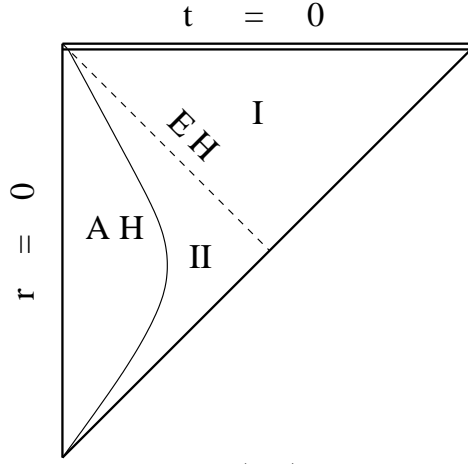,width=0.35\textwidth,angle=0}
 \caption{The Penrose diagram for the solutions given  by
 Eq.(\ref{2.11}) for $0 < k < 1$. The apparent horizon $x = x_{0}$,
 denoted by the  curve $AH$ is
 always timelike. In Region $II$ 
 we have $\theta_{l} > 0$ and $\theta_{l}\theta_{n} < 0$, but in Region $I$
    all the rings of constant $r$ 
  and $t$ are trapped as now we have $\theta_{l} < 0$ and 
  $\theta_{l}\theta_{n} >0$. The spacetime is singular on the double line 
  $t = 0$.   }
 \end{center}
 \end{figure}
%%%%%%%%%%%%%%%%%%%%%%%%%%%%%%%%%%%%%%%%%%%%%%%%%%%%%%%%%%%%%%%%%%%%%%%%%%%%
%%

%%%%%%%%%%%%%%%%%%%%%%%%%%%%%%%%%%%%%%%%%%%%%%%%%%%%%%%%%%%%%%%%%%%%%%%%%%%%
%%%%
%%%%%%%%%%%%

\section{Self-Similar Solutions of the Second Kind}
\lb{secondkind}

\renewcommand{\theequation}{3.\arabic{equation}}
\setcounter{equation}{0}

When $\alpha \neq 1$,   the term that is proportional to $r^{-2}$
has different power-dependence on $r$ from the term that is
proportional to $t^{-2}$, when they are written out in terms of
$r$ and $x$, since $ t=-r^\alpha e^{-\alpha x}$.
Then, it can be shown that the Einstein field equations   in this
case become
 \bqn
 \lb{3.2a}
 y_{,x} - (1+y)\left(\Psi_{,x} - y\right) - y\Phi_{,x} &=& 0,\\
 \lb{3.2b}
 \Phi _{,xx}+\Phi _{,x}\left( \Phi _{,x}-\Psi _{,x}-y-2\right) &=&0,\\
 \lb{3.2c}
 \Psi _{,xx} - \Psi _{,x}\left(\Phi _{,x}-\Psi _{,x} + y + 1-\alpha \right)
 + \left(1 - \alpha\right)y  &=&0,
 \eqn
and
 \bqn
 \lb{3.3}
 \rho &=& \frac{y}{  l^{2}}\left\{\frac{1}{(\alpha
t)^{2}}e^{-2\Phi}\Psi_{,x}
 - \frac{1}{r^{2}}e^{-2\Psi}\Phi_{,x}\right\},\nb\\
 p &=& \frac{1}{  l^{2}}\left\{\frac{1+y}{r^{2}}e^{-2\Psi}\Phi_{,x}
 - \frac{1}{(\alpha t)^{2}}e^{-2\Phi}\left[\left(1 + y\right)\Psi_{,x}
 - \left(1 - \alpha\right)y\right]\right\},
 \eqn
where in writing Eqs.(\ref{3.2b}) - (\ref{3.3}), we had used
Eq.(\ref{3.2a}). From the above equations we can see that now the
Einstein field equations are already sufficient to determinate
completely the metric coefficients $\Phi(x),\; \Psi(x)$ and
$S(x)$. Following Maeda {\em et al} \cite{Ma02}, it can be shown
that the symmetry of the self-similarity of the second kind is
inconsistent with a polytropic equation of the kind,
 \bq
 \lb{3.3a}
 p = k \rho^{\beta},
 \eq
or
 \bq
 \lb{3.3b}
 p = k n^{\beta},\;\;\;
 \rho = m_{b} n + \frac{p}{\beta -1}, 
 \eq
unless $k  =1$, where the constant $m_{b}$ denotes the mean baryon mass, 
and $n(t,r)$  the baryon number density \cite{ST83}. Thus, 
in the following we shall consider only the case $\beta = 1$. Combining it with
Eq.(\ref{3.3}), we find that
  \bqn
  \lb{3.3ca}
  \left[1 + \left(1 + k\right)y\right]\Phi_{,x} &=& 0,\\
  \lb{3.3cb}
  \left[1 + \left(1 + k\right)y\right]\Psi_{,x} - (1-\alpha)y &=& 0.
  \eqn
Since $\alpha \not= 1$, we must have $\Phi_{,x} = 0$. Therefore, {\em for the 
perfect fluid with the equation of state $p = k \rho$ and self-similarity 
of the second kind,  it must move along the  radial timelike geodesics}. To solve 
Eqs.(\ref{3.2a})-(\ref{3.2c}) and (\ref{3.3cb}), let us consider the two cases
$y \not= -1$ and $y = -1$ separately.

{\bf Case A) $\; y \not =-1$}: In this case, from Eq.(\ref{3.2a}) we find that
\bq
\lb{eq1}
\Psi_{,x} = \frac{y_{,x}}{1+y} + y,
\eq
while from Eq.(\ref{3.3cb}) we obtain
\bq
\lb{eq2}
\Psi_{,xx} = \frac{y_{,x}}{1 + \left(1+k\right)y}
\left[\left(1 - \alpha\right) - \left(1 + k\right)\Psi_{,x}\right].
\eq
Inserting the above expressions into Eq.(\ref{3.2c}) we find that it has three
different solutions, 
\bqn
\lb{eq4}
&& (a)\;\; k = 0,\nb\\
&& (b)\;\; y_{,x} = 0,\nb\\
&& (c)\;\; y_{,x} - (\alpha -2)y(1 + y)  =0.
\eqn

When $k = 0$, it can be shown that Eqs.(\ref{3.2a})-(\ref{3.2c}),
(\ref{3.3ca}) and (\ref{3.3cb}) have the general solution,
\bqn
\lb{eq5}
\Phi(x) &=& \Phi_{0},\;\;\;
\Psi(x) = \ln\left|1 + (\alpha -1)e^{\alpha(x_{0} - x)}\right| + \Psi_{0},\nb\\
S(x) &=& S_{0}\left|1 - e^{\alpha(x_{0} - x)}\right|,\;\;\; (k = 0).
\eqn
 
When $y_{,x} = 0$, we find that the general solution is given by
\bqn
\lb{eq6}
\Phi(x) &=& \Phi_{0},\;\; S(x) = S_{0}e^{ax},\nb\\ 
\Psi(x) &=& a x + \Psi_{0},\;\;\; a \equiv - \frac{\alpha}{1+k}.
\eqn

When $y_{,x} - (\alpha -2)y(1 + y)  =0$, it can be shown that the corresponding
solution is given by Eq.(\ref{eq6}) with $\alpha = 2$. Thus, this case is a
particular one of Eq.(\ref{eq6}).

{\bf Case B) $\; y =-1$}: In this case from Eq.(\ref{3.3cb}) we find that
\bq
\lb{eq7}
\Psi_{,x} =  \frac{1-\alpha}{k}.
\eq
Inserting it into Eq.(\ref{3.2c}) we find that there are only two solutions
given, respectively, by
\bqn
\lb{eq8}
& & (i)\;\; k = 1,\nb\\
& & (ii)\;\; \alpha = 1 + k.
\eqn

When $k = 1$, the general solution is given by
\bqn
\lb{eq9}
\Phi(x) &=& \Phi_{0},\;\; S(x) = S_{0}e^{-x},\nb\\ 
\Psi(x) &=& (1-\alpha) x + \Psi_{0},\;\;\;(k = 1),  
\eqn
and when $\alpha = 1 + k$, we have
\bqn
\lb{eq10}
\Phi(x) &=& \Phi_{0},\;\; S(x) = S_{0}e^{-x},\nb\\ 
\Psi(x) &=& - x + \Psi_{0},\;\;\; \alpha = 1 + k. 
\eqn

Thus, the most general solutions with self-similarity of the second kind 
($\alpha \not= 0, 1$) and the equation of state $p = k\rho$ consist of
four classes of solutions, given, respectively, by Eqs.(\ref{eq5}),
(\ref{eq6}), (\ref{eq9}) and (\ref{eq10}). In the following, let us consider
them separately.

\subsection{$y \not= -1,\; k = 0$}

In this case, the solutions are given by Eq.(\ref{eq5}). Applying the  
conditions (\ref{cd1})-(\ref{cd4}) to them, we find that $\Phi_{0} = 0$,
$S_{0} = e^{\Psi_{0}}$ and $\alpha < 1$. Then, using the transformations 
(\ref{rescaling}) we can further set $\Psi_{0} = 0$. Thus,  
the corresponding metric finally  takes the form,
\bq
\lb{eq12}
ds^{2} = l^{2} \left\{dt^{2} 
- \left[1 - (1-\alpha) e^{\alpha(x_{0} -x)}\right]^{2} dr^{2} 
- r^{2}\left[1 -  e^{\alpha(x_{0} -x)}\right]^{2} d\theta^{2}\right\},
\;\;\; (\alpha < 1),
\eq 
and  the energy density and pressure of the fluid are given by
\bqn
\lb{eq13}
\rho &=& \frac{\alpha^{2}(1-\alpha)e^{2\alpha(x_{0} -x)}}
{  l^{2}(\alpha t)^{2}\left|1 -  e^{\alpha(x_{0} -x)}\right|
\left|1 - (1-\alpha) e^{\alpha(x_{0} -x)}\right|},\nb\\
p &=& 0,\;\;\; (\alpha < 1).
\eqn
From the above expressions we can see that the spacetime is singular on the 
hypersurfaces,
\bq
\lb{eq14}
(i)\;\; t = 0,\;\;\;\;
(ii)\;\; x = x_{0},\;\;\;\;
(iii)\;\; x = x_{1},
\eq
where
\bq
\lb{eq15}
x_{1} \equiv x_{0} - \frac{1}{\alpha}\ln\left(\frac{1}{1-\alpha}\right) 
< x_{0}.
\eq
On the other hand, from Eq.(\ref{B.9}) we find that
\bqn
\lb{eq16a}
\theta_{l} &=& \frac{f}{l^{2}r\left|1 - e^{\alpha(x_{0} -x)}\right|}
\left[1 -  \left(\frac{r}{r_{A}}\right)^{1-\alpha}\right],\nb\\
\theta_{n} &=& - \frac{g}{l^{2}r\left|1 -  e^{\alpha(x_{0} -x)}\right|}
\left[1 + \left(\frac{r}{r_{A}}\right)^{1-\alpha}\right],\;\; (0< \alpha < 1),
\eqn
for $0 < \alpha < 1$, and 
\bqn
\lb{eq16b}
\theta_{l} &=& \frac{f}{l^{2}r\left|1 - e^{\alpha(x_{0} -x)}\right|}
\left[1 + \left(\frac{r}{r_{A}}\right)^{1-\alpha}\right],\nb\\
\theta_{n} &=& - \frac{g}{l^{2}r\left|1 - e^{\alpha(x_{0} -x)}\right|}
\left[1 - \left(\frac{r}{r_{A}}\right)^{1-\alpha}\right],\;\; (\alpha < 0),
\eqn 
for $\alpha < 0$, where $r_{A} \equiv e^{-\alpha x_{0}/(1-\alpha)}$.

From Eq.(\ref{eq16a}) we can see that, in the case $0 < \alpha < 1$,
$\theta_{n}$ is always negative for any $r \in [0, \infty)$, but 
$\theta_{l}$ is negative when $ r > r_{A}$, zero when $ r = r_{A}$
and positive when $ r < r_{A}$ [cf. Fig. 5]. That is, the spacetime is
closed far away from the axis even at the very beginning ($t = -\infty$). 
This property makes the solution very difficult to be considered as 
representing gravitational collapse.  

When $\alpha < 0$, from Eq.(\ref{eq16b}) we can see that $\theta_{l}$ now is
always positive, but $\theta_{n}$ changes signs at $r = r_{A}$. In particular,
it shows that the ingoing radial null geodesics become expanding in the
region $r > r_{A}$. With this odd property, it is also very  difficult to   
consider the corresponding solutions as representing 
gravitational collapse.

%%%%%%%%%%%%%%%%%%%%%%%%%%%%%%%%%%%%%%%%%%%%%%%%%%%%%%%%%%%%%%%%%%%%%%%%%%%%
%%%%%%%
 \begin{figure}[htbp]
 \begin{center}
 \label{fig5}
 \leavevmode
  \epsfig{file=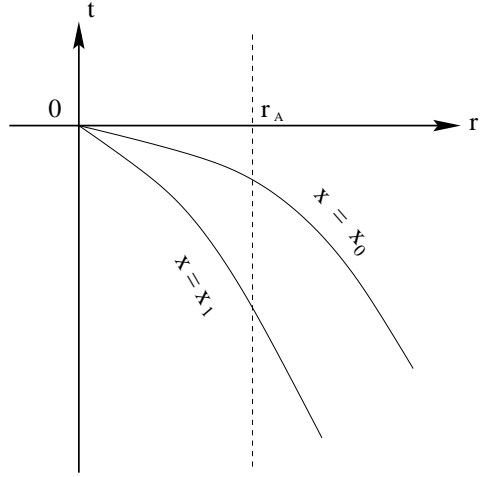,width=0.35\textwidth,angle=0}
 \caption{The  spacetime in the ($t, r$)-plane for the solutions given 
 by the metric  (\ref{eq12}). It is singular on the hypersurfaces
 $t = 0$, $\; x= x_{0}$ and  $ x= x_{1}$, where $x_{0} > x_{1}$.
   }
 \end{center}
 \end{figure}

%%%%%%%%%%%%%%%%%%%%%%%%%%%%%%%%%%%%%%%%%%%%%%%%%%%%%%%%%%%%%%%%%%%%%%%%%%%%
%%%%%%%

\subsection{$y = Const. (\not= -1)$}

In this case the solutions are given by Eq.(\ref{eq6}). It can be shown that 
the regularity conditions (\ref{cd1})-(\ref{cd3}) and the gauge one
(\ref{cd4}) require $\Phi_{0} = 0$, $S_{0} = e^{\Psi_{0}}/(1 + a)$
and $\alpha < 1 + k$. On the other hand, using the transformations 
(\ref{rescaling}), we can further set $\Psi_{0} = 0$. Then, we find that  
\bq
\lb{eq17}
ds^{2} = l^{2}\left\{dt^{2} - e^{2ax}\left[dr^{2}
+ \left(\frac{r}{1+a}\right)^{2}d\theta^{2}\right]\right\},
\;\; (\alpha < 1 + k),  
\eq
and the corresponding physical quantities are  
\bqn
\lb{eq17a}
p &=& k\rho = \frac{k\alpha^{2}}
{  l^{2}(1+k)^{2}(\alpha t)^{2}},\nb\\
\theta_{l} &=& \frac{f e^{-ax}}{l^{2}(1+k)r}
\left[(1 + k)(1+a) 
- \frac{r^{1+a}}{(-t)^{k/(1+k)}}\right],\nb\\
\theta_{n} &=& - \frac{g e^{-ax}}{l^{2}(1+k)r}
\left[(1 + k)(1+a) 
+ \frac{r^{1+a}}{(-t)^{k/(1+k)}}\right].
\eqn
From the above we can see that the spacetime is singular on the hypersurface
$ t = 0$, and the singularity is not naked. In fact, it is covered by the 
apparent horizon located on the hypersurface,
\bq
\lb{eq18}
r_{AH}(t) = \left[(1 + k)(1+a)(-t)^{k/(1+k)}\right]^{1/(1+a)}.
\eq
The normal vector to the hypersurface $r - r_{AH}(t) = 0$ is given by $n_{\alpha}
\equiv \partial(r - r_{AH}(t))/\partial x^{\alpha}$. Then, we find that 
\bq
\lb{eq19}
n_{\alpha}n^{\alpha} = -l^{-2}\left(1+k -\alpha\right)^{2\alpha/
(1+k -\alpha)}
(-t)^{-2(1-\alpha)/(1+k -\alpha)}\left(1 - k^{2}\right).
\eq
Thus, the apparent horizon is timelike for $0 \le k < 1$, 
and the corresponding Penrose diagram is that of Fig. 4. 
When $k = 1$, it is null, and the corresponding
Penrose diagram is that of Fig. 2.

\subsection{$y = -1, \; k = 1$}

In this case the solutions are given by Eq.(\ref{eq9}). The transformations 
(\ref{rescaling}) and the gauge condition (\ref{cd4}) enable us to set
$\Phi_{0} = \Psi_{0} = 0,\; S_{0} = 1$. Then, the metric  takes the form
\bq
\lb{eq19a}
ds^{2} = l^{2}\left\{dt^{2} - e^{2(1-\alpha)x}dr^{2}
- (-t)^{2/\alpha}d\theta^{2}\right\},
\eq
from which we can see that the solutions are Kantowski-Sachs like and do not
satisfy the regular  conditions (\ref{cd1}) and (\ref{cd2}).
The corresponding energy density is given by
\bq
\lb{eq20}
\rho = p = \frac{\alpha -1}{  l^{2}(\alpha t)^{2}}.
\eq
Thus, to have the energy density be positive, we must assume that $\alpha > 1$.

\subsection{$y = -1,\; \alpha = 1+ k$}

In this case, the solutions are given by Eq.(\ref{eq10}). Using the transformations
(\ref{rescaling}) and the gauge condition (\ref{cd4}) we can set
$\Phi_{0} = 0$ and $S_{0} = 1$. Then, the metric takes the form,
\bq
\lb{eq21}
ds^{2} = l^{2}\left\{dt^{2} - e^{2(x_{0}-x)}dr^{2}
- (-t)^{2/\alpha}d\theta^{2}\right\},
\eq
where $\alpha = 1 + k$ and $x_{0} \equiv \Psi_{0}$. 
Thus, in this case the solutions are also 
 Kantowski-Sachs like, and the corresponding energy density and pressure of the
 fluid are given by 
\bq
\lb{eq22}
p = k \rho = \frac{k}{  l^{2}(\alpha t)^{2}}.
\eq

As we mentioned in the last section, when $k = 1$ the perfect fluid is energetically 
equivalent to a massless scalar field. Indeed, the solutions given by
metric (\ref{eq19a}) and the ones given by
 Eqs.(\ref{eq17}) and (\ref{eq21}) with $k = 1$
are the solutions found lately for a massless scalar field in \cite{Fatima2}.   
%%%%%%%%%%%%%%%%%%%%%%%%%%%%%%%%%%%%%%%%%%%%%%%%%%%%%%%%%%%%%%%%%%%%%%%

\section{Self-Similar Solutions of the Zeroth Kind}
\lb{zerothkind}

\renewcommand{\theequation}{4.\arabic{equation}}
\setcounter{equation}{0}

In this case, combining Eqs.(\ref{1.3}) and (\ref{A.8}), we find
that the Einstein field equations can be cast in the forms,
 \bqn
 \lb{4.1a}
 y_{,x} - (1+y)\left(\Psi_{,x} - y\right) - y\Phi_{,x} &=& 0,\\
 \lb{4.1b}
 \Phi _{,xx} + \Phi _{,x}\left(\Phi _{,x}-\Psi _{,x}-y-2\right) &=&0,\\
 \lb{4.1c}
 \Psi _{,xx} - \Psi _{,x}\left(\Phi _{,x}-\Psi _{,x} + y + 1\right)
 +  y  &=&0,
 \eqn
and
 \bqn
 \lb{4.2}
 \rho &=& \frac{y}{  l^{2} r^{2}}\left\{r^{2}e^{-2\Phi}\Psi_{,x}
          -  e^{-2\Psi}\Phi_{,x}\right\},\nb\\
 p &=& \frac{1}{  l^{2} r^{2}}\left\{(1+y)e^{-2\Psi}\Phi_{,x}
      - r^{2}e^{-2\Phi}\left[\left(1 + y\right)\Psi_{,x}
      -y\right]\right\},
  \eqn
where in writing Eqs.(\ref{4.1b}) - (\ref{4.2}), we had used
Eq.(\ref{4.1a}). Similar to the solutions with self-similarity of
the second kind, now the Einstein field equations already
determinate completely the metric coefficients. It can be also
shown that {\em the self-similarity of the zeroth kind is
inconsistent with the equation of state given by Eq.(\ref{3.3a}), 
unless $\beta = 1$}. Then, 
following the last section, let us  consider  solutions that 
satisfy the equation of state (\ref{2.4}), which, together with  
Eq.(\ref{4.2}), yields,
\bqn
\lb{4.2aa}
\left[1 + \left(1 + k\right) y\right] \Phi,_{x} &=& 0, \\
\lb{4.2ab}
\left[1 + \left(1 + k\right) y\right] \Psi,_{x} - y &=& 0.
\eqn
The above   equations have the solution 
\bq
\lb{4.2ac}
 \Phi(x) = \Phi_{0}, \;\;\;
 \Psi_{,x} = \frac{y}{1 + \left(1 + k\right) y}.
\eq
Thus, similar to the case of self-similarity of the second kind,
now {\em the fluid must also move  along radial timelike geodesics}.
Inserting the above expressions into Eqs.(\ref{4.1a})-(\ref{4.1c}), we find
that   solutions exist only for three special cases, $k = 0,\; \pm 1$.

\subsection {$k = 0$}

When $k = 0$,   the general solution is given by,
 \bqn
 \lb{4.11}
 \Phi(x) &=& \Phi_{0},\;\;\; S(x) = S_{0}\left(x_{0} - x\right),\nb\\
 \Psi(x) &=& \ln\left(x_{0} - x -1\right) + \Psi_{0},\;\; (k = 0).
 \eqn
 Then, it can be shown that the   conditions (\ref{cd1})-(\ref{cd4}) require
 $\Phi_{0} = 0$ and $S_{0} = e^{\Psi_{0}}$. On the other hand, using the 
 transformations (\ref{rescalingb}), we can further set $\Psi_{0} = 0$.
 Then, the solutions are finally given by 
 \bq
 \lb{4.11a}
 ds^{2} = l^{2}\left\{dt^{2} - \left(x_{0} - x - 1\right)^{2}dr^{2}
 - r^{2}(x_{0} - x)^{2}d\theta^{2}\right\},
 \eq
 and  
the corresponding energy density is given by
 \bq
 \lb{4.12}
 \rho = \frac{1}{  l^{2}\left(x_{0} - x -1\right)
 \left(x_{0} - x \right)},\;\;\; p = 0.
 \eq
From this expression  we can see that the 
spacetime is singular on the hypersurfaces $x = x_{0} -1$ and $x =
x_{0}$ [cf. Fig. 6]. From Eq.(\ref{B.8}) we also find that
\bqn
\lb{4.11b}
\theta_{l} &=& \frac{f}{l^{2}(x_{0} - x)r}\left( 1 + r\right),\nb\\
\theta_{n} &=& - \frac{g}{l^{2}(x_{0} - x)r}\left( 1 - r\right).
\eqn
Thus, $\theta_{l}$ is  positive for $x < x_{0}$ and negative for $x > x_{0}$,
while the signs of $\theta_{n}$ get changed when across the hypersurfaces
$r =1$ and $x = x_{0}$.  With this property, one can see
that the solutions cannot be considered as representing gravitational
collapse.

%%%%%%%%%%%%%%%%%%%%%%%%%%%%%%%%%%%%%%%%%%%%%%%%%%%%%%%%%%%%%%%%%%%%%%%%%%%%
%%%%%%%
 \begin{figure}[htbp]
 \begin{center}
 \label{figx}
 \leavevmode
  \epsfig{file=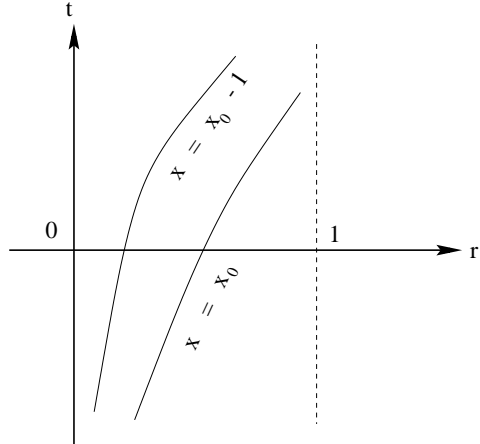,width=0.35\textwidth,angle=0}
 \caption{The  spacetime in the ($t, r$)-plane for the solutions given by
 Eq.(\ref{4.11a}). It is singular on the hypersurfaces $x = x_{0}$ and
 $x = x_{0} - 1$.  The signs of $\theta_{n}$ change when across the hypersurfaces
$r =1$ and $x = x_{0}$. }
 \end{center}
 \end{figure}

%%%%%%%%%%%%%%%%%%%%%%%%%%%%%%%%%%%%%%%%%%%%%%%%%%%%%%%%%%%%%%%%%%%%%%%%%%%%
%%%%%%%

\subsection {$k = -1$}

When $k = -1$, the general solution is given by 
  \bq
  \lb{4.16}
  \Phi(x) = \Phi_{0},\;\;\;
  \Psi(x) = ax + \Psi_{0},\;\;\;
  S(x) = S_{0}e^{ax},
  \eq
 for which we have
  \bq
  \lb{4.17}
  \rho = -p =  \frac{a^{2}e^{-2\Phi_{0}}}{  l^{2}}.
  \eq
 This is a 3-dimensional de-Sitter   solution. In fact, setting 
\bqn
\lb{4.17a}
t &=& e^{-\Phi_{0}} \bar{t} + \frac{\Psi_{0}}{a},\;\;\;\;
r = \left[\left(1 + a\right)\bar{r}\right]^{1/(1+a)},\nb\\
\theta &=& \frac{e^{\Psi_{0}}}{(1+a)S_{0}}\bar{\theta},\;\;\;\;
\beta \equiv a e^{-\Phi_{0}},
\eqn
we find that the corresponding metric can be cast in the form
  \bq
  \lb{4.18}
  ds^{2} =l^{2}\left\{d\bar{t}^{2} - e^{-2\beta \bar{t}}\left(d\bar{r}^{2} +
  \bar{r}^{2}d\bar{\theta}^{2}\right)\right\}.
  \eq

\subsection {$k = 1$}

In this case, the general solution is given by 
  \bq
  \lb{4.16aa}
  \Phi(x) = \Phi_{0},\;\;\;
  \Psi(x) = x + \Psi_{0},\;\;\;
  S(x) = S_{0}e^{-x},
  \eq
 for which we have
  \bq
  \lb{4.17aa}
  \rho = p = - \frac{e^{-2\Phi_{0}}}{  l^{2}}.
  \eq
Clearly, now none of the three energy conditions is satisfied, and the physics of
the solution is unclear (if there is any).

\section{Conclusions}

In this paper we have first generalized the notion of kinematic
self-similarity of four-dimensional spacetimes to any dimensions
for the metric given by Eq.(\ref{1.1}), and then restricted
ourselves to $(2+1)$-dimensional spacetimes with circular symmetry.

In Sec. $II$, we have studied solutions of the Einstein field
equations with homothetic (the first kind) self-similarity for a
perfect fluid. It has been shown that the only    equation of
state  that takes the form $p = p(\rho)$ and is consistent with
the self-similarity is $p = k \rho$,
where $k$ is a constant. In the latter case,   a master
equation has been found,  Eq.(\ref{2.6c}). Then, the general
solutions for the stiff fluid ($k = 1$) have been given in
closed form. It has been found that some of these solutions
represent formation of marginally naked singularities and the
others represent the formation of black holes from gravitational
collapse of the self-similar fluid. All the solutions of dust
fluid ($k = 0$) have also been given, and found that they are
all Kantowski-Sachas-like. When $k \not=0, 1$, only
particular solutions have been found and shown that the
corresponding collapse always forms black holes.

In Secs. $III$ and $IV$, all the solutions with self-similarity of
the second or zeroth kind and the equation of state $p = k
\rho$ have been found and studied. In particular, it has been 
shown that the fluid must move along timelike radial geodesics. 
It has been also shown that some of
those solutions represent gravitational collapse of the perfect
fluid, while the others don't. In the collapsing case,  
black holes are always formed.

In critical Type II collapse, all the critical solutions found so
far have either discrete self-similarity or kinematic self-similarity
of the first kind, and no critical solutions with kinematic self-similarity
of the zeroth or second kind have been found, yet \cite{Gun00,Wang01}. 
However, in Statistical Mechanics, critical solutions with kinematic self-similarity
of the   second kind seem more generic than those of the first kind \cite{Baren}.
Thus, it would be very interesting to study the linear perturbations of
the solutions found in this paper with  self-similarity
of the zeroth or second kind.

Finally, we would like to note that Ida recently showed that a (2+1)-dimensional 
gravity theory which satisfies the dominant energy condition forbids 
the existence of black holes \cite{Ida00}. This result does not 
contradict with the  ones obtained here, as in our case the surfaces 
of apparent horizons of the black holes are not compact, a precondition 
that was assumed in \cite{Ida00}.

\section*{Append A:    Circularly Symmetric Spacetimes with Kinematic
 Self-Similarity }

\lb{appendixa}
\renewcommand{\theequation}{A.\arabic{equation}}
\setcounter{equation}{0}

The general metric of $\left( 2+1\right) $-dimensional  spacetimes
with circular symmetry can be cast in the form,
 \bq
 \lb{A.1}
 ds^{2}= l^{2}\left\{e^{2\Phi\left( t,r\right)}dt^{2}
        -e^{2\Psi\left( t,r\right)}dr^{2}
    -r^{2}S^{2}\left(t,r\right)d\theta ^{2}\right\},
 \eq
where $l$ is a constant and has  dimension of length. Then, it is
easy to show that the coordinates $\{x^{\mu}\} = \{t, \; r, \;
\theta\}$, the Christoffel symbols, $\Gamma^{\lambda}_{\mu\nu}$,
the Riemann tensor, $R^{\sigma}_{\mu\nu\lambda}$, the Ricci
tensor, $R_{\mu\nu}$, and the Einstein tensor, $G_{\mu\nu}$, are
all {\em dimensionless}, while the Ricci scalar, $R$, has the
dimension of $l^{-2}$, and the Kretschmann scalar, ${\cal{R}}
\equiv R^{\sigma\mu\nu\lambda} R_{\sigma\mu\nu\lambda}$, has the
dimension of $l^{-4}$.

For the metric (\ref{A.1}), we find that the non-vanishing
Christoffel symbols  are given by \bqn \lb{A.2} \Gamma _{00}^{0}
&=&\Phi _{,t},\;\;\; \Gamma_{01}^{0}=\Phi _{,r},\;\;\;
\Gamma_{11}^{0}= e^{2\left( \Psi -\Phi \right)}\Psi _{,t}, \;\;\;
\Gamma_{22}^{0} =  r^{2}e^{-2\Phi }SS_{,t},\nb\\
\Gamma _{00}^{1} &=&e^{2\left( \Phi -\Psi \right)}\Phi
_{,r},\;\;\; \Gamma _{01}^{1}=\Psi _{,t}, \;\;\; \Gamma _{11}^{1}
=\Psi _{,r},\;\;\;
\Gamma _{22}^{1}= re^{-2\Psi }S \left( S+rS_{,r}\right), \nonumber \\
\Gamma _{02}^{2} &=&\frac{S_{,t}}{S},\;\;\; \Gamma
_{12}^{2}=\frac{S+rS_{,r}}{rS}, \eqn
 and the Einstein tensor has the following non-zero components
 \bqn
 \lb{A.3}
  G_{tt} &=&\frac{e^{-2\Psi}}{rS}\left\{re^{2\Psi} S_{,t}\Psi_{,t} -
e^{2\Phi}\left[rS_{,rr} + 2S_{,r}
- \left(S + rS_{,r}\right)\Psi_{,r}\right]\right\}, \nb\\
G_{tr} &=& - \frac {1}{rS}\left[rS_{,tr} - S_{,t}\left(r\Phi_{,r}
-1\right)
- \left(rS_{,r} + S\right)\Psi_{,t}\right], \nb\\
G_{rr} &=&\frac {e^{-2\Phi}}{rS}\left[e^{2\Phi}\left(rS_{,r} +
S\right)\Phi_{,r}
-  re^{2\Psi}\left(S_{,tt} -  \Phi _{,t} S_{,t}\right)\right],  \nb\\
G_{\theta \theta } &=&-r^2S^2\left\{ e^{-2\Phi}\left[\Psi_{,tt} +
\left(\Psi_{,t}
   - \Phi_{,t}\right)\Psi_{,t}\right]\right.\nb\\
& & - \left. e^{-2\Psi}\left[\Phi_{,rr} + \left(\Phi_{,r} -
\Psi_{,r}\right)\Phi_{,r}\right]\right\}. \eqn

\subsection*{A.1 $\;$ Spacetimes with Self-Similarity of the Zeroth Kind}

To study solutions with self-similarity of the zeroth kind, let us
first introduce the self-similar variables, $x$ and $\tau$ via the
relations,
 \bq
 \lb{A.4a}
 x = \ln (r) - t,\;\;\;\;
 \tau =t,
 \eq
 or inversely
 \bq
 \lb{A.4b}
 t= \tau ,\;\;\;\;
 r= e^{x+\tau}.
 \eq
Then, for any given function $f(t,r)$   we have
 \bqn
 \lb{A.5}
 f_{,t} &=&f_{,\tau }-f_{,x},\;\;\;\;\;\;\;\;f_{,r}=\frac {1}{r}f_{,x},
\nb\\
 f_{,tr} &=&-\frac {1}{r}\left(f_{,xx}-f_{,\tau x}\right) ,\;\;\;
 f_{,rr}=\frac {1}{r^2}\left(f_{,xx}-f_{,x}\right),  \nonumber \\
 f_{,tt} &=&f_{,\tau \tau }-2f_{,\tau x}+f_{,xx}.
 \eqn
 Substituting these equations into Eq.(\ref{A.3}), we find that
 \bqn
 \lb{A.6}
 G_{tt} &=&-\frac{e^{-2\Psi }}{r^2S}\left\{ e^{2\Phi }\left[
            S_{,xx}+S_{,x}-\Psi _{,x}\left( S_{,x}+S\right) \right] \right.
\nb\\
        & & \left. -r^2e^{2\Psi }\left( S_{,\tau }-S_{,x}\right)
             \left(\Psi_{,\tau}-\Psi _{,x}\right) \right\} ,  \nb\\
 G_{tr} &=&\frac 1{rS}\left[ S_{,xx}-S_{,\tau x}+\left( S_{,x}+S\right)
           \left( \Psi _{,\tau }-\Psi _{,x}\right) \right.\nb\\
        & & \left. +\left( S_{,\tau }-S_{,x}\right)
           \left( \Phi _{,x}-1\right) \right] ,  \nb\\
 G_{rr} &=&\frac{e^{-2\Phi }}{r^2S}\left\{ e^{2\Phi }\Phi_{,x}
            \left(S_{,x}+S\right) -r^2e^{2\Psi }\left[ S_{,\tau \tau }
            -2S_{,\tau x}+S_{,xx}\right. \right.  \nonumber \\
        & &\left. \left. +\left( S_{,\tau }-S_{,x}\right)
        \left( \Phi _{,x}-\Phi_{,\tau }\right) \right] \right\} ,  \nb\\
G_{\theta \theta } &=&S^2e^{-2(\Phi +\Psi )}\left\{ e^{2\Phi }
                      \left[ \Phi_{,xx}+\Phi _{,x}\left( \Phi _{,x}-\Psi
_{,x}-1\right)
                      \right] \right.\nonumber \\
     & & \left. -r^2e^{2\Psi }\left[ \Psi _{,xx}+\Psi _{,\tau \tau }
         -2\Psi _{,\tau x}\right. \right.  \nonumber \\
     & & \left. \left. -\left( \Psi _{,\tau }-\Psi _{,x}\right)
        \left( \Phi _{,\tau}-\Psi _{,\tau }-\Phi _{,x}+\Psi _{,x}\right)
\right]
        \right\} .
\eqn For the self-similar solutions, the metric coefficients $\Phi
,\;\Psi $ and $S$ are functions of $x$ only.
% \bq
%\lb{A.7}
%\Phi (\tau,x)=\Phi (x),\;\;\;\;
% \Psi (\tau ,x)=\Phi (x),\;\;\;\;
% S(\tau,x)=S(x).
%\eq
 Then, Eq.(\ref{A.6}) reduce to,
 \bqn
 \lb{A.8}
 G_{tt} &=& -\frac{e^{-2\Psi }}{r^2}\left\{ e^{2\Phi }
           \left[y_{,x}+\left( y+1\right) \left( y-\Psi _{,x}\right)\right]
            -r^2e^{2\Psi }\Psi_{,x}y\right\} ,  \nb \\
 G_{tr} &=&\frac 1r\left[ y_{,x}+\left( y+1\right) \left(y -
           \Psi _{,x}\right)-y\Phi _{,x}\right] ,  \nb \\
G_{rr} &=&\frac{e^{-2\Phi }}{r^2}\left\{ e^{2\Phi }\Phi _{,x}
           \left(y+1\right) -r^2e^{2\Psi }\left[ y_{,x}
           +y\left( y-\Phi _{,x}\right) \right]\right\} ,  \nb \\
 G_{\theta \theta } &=&S^2e^{-2(\Phi +\Psi )}\left\{ e^{2\Phi }
       \left[ \Phi_{,xx}+\Phi _{,x}\left( \Phi _{,x}-\Psi _{,x}-1\right)
        \right] \right.\nonumber \\
     && \left. -r^2e^{2\Psi }\left[ \Psi _{,xx}-\Psi _{,x}
     \left( \Phi _{,x}-\Psi_{,x}\right) \right] \right\},
 \eqn
 where
 \bq
 \lb{A.9}
 y \equiv \frac{S_{,x}}S.
 \eq

 \subsection*{A.2 $\;$ Spacetimes with Self-Similarity of the First
 and Second  Kinds}

 To study these kinds of self-similar solutions, let us introduce the
self-similar
 variables, $x$ and $\tau$ by
  \bq
  \lb{A.9aa}
    x= \ln \left(\frac r{\left(-t\right) ^{\frac 1\alpha }}\right)
,\;\;\;\;\;
  \tau =-\ln \left(-t\right),
  \eq
  or inversely,
   \bq
   \lb{A.9a}
   r = e^{(\alpha x - \tau)/\alpha},\;\;\;
   t = - e^{-\tau},
   \eq
where $\alpha$ is a {\it dimensionless} constant. When $\alpha =
1$, the corresponding spacetimes are said to have self-similarity
of the first kind or homothetic self-similarity. Otherwise, they
  are said to have  self-similarity of the second kind.

  For any given function $f\left(t,r\right) $, now we have
  \bqn
  \lb{A.10}
   f_{,t} &=&-\frac 1{\alpha t}\left(\alpha
f_{,\tau }+f_{,x}\right),\;\;\;\;
   f_{,r}=\frac 1rf_{,x},  \nonumber \\
   f_{,tr} &=&-\frac 1{\alpha tr}\left( \alpha f_{,\tau
x}+f_{,xx}\right),\;\;\;\;\;
   f_{,rr}=\frac 1{r^2}\left( f_{,xx}-f_{,x}\right) ,  \nonumber \\
 f_{,tt} &=&\frac 1{\alpha ^2t^2}\left( \alpha ^2f_{,\tau \tau }
            +2\alpha f_{,\tau x}+f_{,xx}+\alpha ^2f_{,\tau }+\alpha
f_{,x}\right) .
 \eqn
Substituting these equations into Eq.(\ref{A.3}), we find that
 \bqn
 \lb{A.11}
 G_{tt} &=&-\frac {1}{\alpha ^2r^2Se^{2\Psi }}\left\{\alpha^2e^{2\Phi}
          \left[S_{,xx}+S_{,x}-\Psi _{,x}\left( S_{,x}+S\right) \right]
\right.
\nb\\
        & &\left. -\frac{r^2}{t^2}\Psi _{,x}S_{,x}e^{2\Psi }\right.
\nonumber
\\
        & &\left. -\frac{\alpha r^2}{t^2}e^{2\Psi }\left( \alpha \Psi_{,\tau
}S_{,\tau }
        +\Psi _{,x}S_{,\tau }+\Psi _{,\tau}S_{,x}\right) \right\},\nb\\
 G_{tr} &=&\frac 1{\alpha trS}\left\{ S_{,xx}-\Psi
_{,x}\left(S_{,x}+S\right)
           -S_{,x}\left( \Phi _{,x}-1\right) \right.  \nonumber \\
        & &\left. +\alpha \left[ S_{,\tau x}-\Psi _{,\tau }\left(
S_{,x}+S\right)
           -S_{,\tau }\left( \Phi _{,x}-1\right) \right] \right\},
           \nb\\
 G_{rr} &=&\frac 1{\alpha ^2r^2Se^{2\Phi }}\left\{ \alpha^2e^{2\Phi }
            \left[\Phi _{,x}\left( S_{,x}+S\right) \right] \right.
\nonumber \\
         & &\left. -\frac{r^2}{t^2}e^{2\Psi }\left( S_{,xx}-S_{,x}\Phi
_{,x}\right)
            \right.  \nonumber \\
        & & \left. -\frac{\alpha r^2}{t^2}e^{2\Psi }\left\{ \alpha
            S_{,\tau \tau}+2S_{,\tau x}-S_{,x}\left( \Phi _{,\tau }-1\right)
            \right. \right.\nonumber \\
        & & \left. \left. -S_{,\tau }\left[ \alpha \left( \Phi _{,\tau
}-1\right)
            +\Phi _{,x}\right] \right\} \right\},  \nb \\
 G_{\theta \theta } &=&\frac{S^2}{\alpha ^2}\left\{ \alpha ^2e^{-2\Psi}
          \left[ \Phi _{,xx}+\Phi _{,x}\left( \Phi _{,x}-\Psi _{,x}-1\right)
\right]
           \right.  \nonumber \\
      & &  \left. -\frac{r^2}{t^2}e^{-2\Phi }\left[ \Psi _{,xx}-\Psi _{,x}
          \left( \Phi_{,x}-\Psi _{,x}-\alpha \right) \right] \right.
\nonumber
\\
      &&  \left. -\frac{\alpha r^2}{t^2}e^{-2\Phi }\left\{ \alpha
          \Psi _{,\tau \tau}+2\Psi _{,\tau x}-\Psi _{,\tau }\left[ \alpha
          \left( \Phi _{,\tau }-\Psi_{,\tau }-1\right) +\Phi _{,x}-\Psi
_{,x}\right]
          \right. \right.  \nonumber\\
      && \left. \left. -\Psi _{,x}\left( \Phi _{,\tau }-\Psi
_{,\tau }\right)
          \right\} \right\}.
 \eqn

 For  the self-similar solutions, the metric coefficients are also
functions of $x$
 only, but now with $x$ being given by Eq.(\ref{1.10}). Thus, for the
 self-similar solutions,   Eq.(\ref{A.11}) reduces to
  \bqn
  \lb{A.12}
  G_{tt} &=&\frac 1{\alpha ^2t^2}\Psi
_{,x}y-\frac{1}{r^2}e^{2\left(\Phi -\Psi \right)}
            \left[ y_{,x}+\left(y+1\right) \left( y-\Psi _{,x}\right)
\right],
   \nb \\
 G_{tr} &=&\frac 1{\alpha tr}\left[ y_{,x}+\left( y+1\right)
           \left( y-\Psi_{,x}\right) -y\Phi _{,x}\right] ,  \nb \\
 G_{rr} &=&-\frac 1{\alpha ^2t^2}e^{2\left( \Psi -\Phi \right)}
           \left[y_{,x}+y\left( y-\Phi _{,x}+\alpha \right) \right]
           +\frac 1{r^2}\Phi_{,x}\left( y+1\right) ,  \nb \\
 G_{\theta \theta } &=&S^2\left\{ e^{-2\Psi }\left[ \Phi _{,xx}
    +\Phi_{,x}\left( \Phi _{,x}-\Psi _{,x}-1\right) \right] \right.
\nonumber
\\
  &&\left. -\frac{r^2}{\alpha ^2t^2}e^{-2\Phi }\left[ \Psi _{,xx}-\Psi_{,x}
     \left( \Phi _{,x}-\Psi _{,x}-\alpha \right) \right] \right\},
 \eqn
where $y$ is given by Eq.(\ref{A.9}).

\section*{{\bf Append B}:  Apparent Horizons of Circularly Symmetric
Spacetimes} 

\lb{appendix}
\renewcommand{\theequation}{B.\arabic{equation}} \setcounter{equation}{0}

In \cite{HW02}, the ingoing and outgoing radial null geodesics
were studied in double null coordinates, and apparent horizon was
defined as the outmost hypersurface where the expansion of outgoing null
geodesics vanishes. In this appendix, we shall write down the
expansions in terms of the self-similar variables. To do this, let
us first introduce two null coordinates $u$ and $v$ via the
relations
  \bq
  \lb{B.1}
  du = f\left(e^{\Phi} dt - e^{\Psi}dr\right),\;\;\;
  dv = g\left(e^{\Phi} dt + e^{\Psi}dr\right),
  \eq
 where $f$ and $g$ satisfy the integrability conditions for $u$
and $v$,
\bq
\lb{B.1aa}
\frac{\partial^{2} u}{\partial t\partial r} = 
\frac{\partial^{2} u}{\partial r\partial t},\;\;\;\;
\frac{\partial^{2} v}{\partial t\partial r} = 
\frac{\partial^{2} v}{\partial r\partial t}.
\eq
Without loss of generality, we shall assume that they are
all strictly positive,
 \bq
 \lb{B.1a}
 f > 0,\;\;\;\;
 g > 0.
 \eq
From the above we can see that the rays moving along the hypersurfaces
$u = Const.$ are outgoing, while the ones moving along the hypersurfaces
$v = Const.$ are ingoing.
Then, it is easy to show that, in terms of $u$ and $v$,
 the metric (\ref{A.1}) takes the form
  \bq
  \lb{B.2}
  ds^{2} = l^{2}\left\{2 e^{2\sigma(u, v)} dudv -
R^{2}(u,v)d\theta^{2}\right\},
  \eq
 where
  \bq
  \lb{B.3}
  \sigma(u,v) = - \frac{1}{2}\ln\left(2fg\right),\;\;\;
  R(u,v) = rS.
  \eq

On the other hand, from Eq.(\ref{B.1}) we find that
  \bqn
  \lb{B.4}
  \frac{\partial t}{\partial u} &=& \frac{1}{2f}e^{-\Phi},\;\;\;
  \frac{\partial t}{\partial v} = \frac{1}{2g}e^{-\Phi},\nb\\
  \frac{\partial r}{\partial u} &=& - \frac{1}{2f}e^{-\Psi},\;\;\;
  \frac{\partial r}{\partial v} = \frac{1}{2g}e^{-\Psi}.
  \eqn
Then, the expansions of the outgoing and ingoing null geodesics are defined as
\cite{HW02},
  \bqn
  \lb{B.5}
  \theta_{l} &\equiv& \nabla_{\lambda}{l^{\lambda}}
  = e^{-2\sigma}\frac{R_{,v}}{l^{2}R}
  = \frac{f}{l^{2}R}\left(e^{-\Phi}R_{,t} + e^{-\Psi}R_{,r}\right),\nb\\
  \theta_{n} &\equiv& \nabla_{\lambda}{n^{\lambda}}
  = e^{-2\sigma}\frac{R_{,u}}{l^{2}R}
  = \frac{g}{l^{2}R}\left(e^{-\Phi}R_{,t} - e^{-\Psi}R_{,r}\right),
  \eqn
where $\nabla_{\lambda}$ denotes the covariant derivative, and
$l^{\lambda}$ ($n^{\lambda}$) is the null vector defined the outgoing (ingoing) null
geodesics and given by
 \bq
 \lb{B.6}
 l_{\lambda} \equiv \frac{\partial u}{\partial x^{\lambda}} =
 \delta^{u}_{\lambda},\;\;\;
 n_{\lambda} \equiv \frac{\partial v}{\partial x^{\lambda}} =
 \delta^{v}_{\lambda}.
 \eq

For the self-similar solutions of the zeroth kind, Eq.(\ref{B.5}) takes
the form,
 \bqn
 \lb{B.8}
 \theta_{l} &=& \frac{f}{l^{2}r}\left\{\left(1+y\right)e^{-\Psi}
              - y e^{x + \tau - \Phi}\right\},\nb\\
 \theta_{n} &=& - \frac{g}{l^{2}r}\left\{\left(1+y\right)e^{-\Psi}
              + y e^{x + \tau - \Phi}\right\}, \;\; (\alpha = 0),
 \eqn
while for the ones of the first or second kind, it becomes
 \bqn
 \lb{B.9}
 \theta_{l} &=& \frac{f}{\alpha l^{2}r}\left\{\alpha\left(1+y\right)e^{-\Psi}
              + y e^{x + (\alpha -1)\tau/\alpha - \Phi}\right\},\nb\\
  \theta_{n} &=& - \frac{g}{\alpha l^{2}r}\left\{\alpha\left(1+y\right)e^{-\Psi}
              - y e^{x + (\alpha -1)\tau/\alpha - \Phi}\right\},
              \;\;(\alpha \not= 0).
 \eqn
 
 Setting $\theta_{l} = 0$ in the above expressions, we shall find the
location of the apparent horizons.

%%%%%%%%%%%%%%%%%%%%%%%%%%%%%%%%%%%
%\newpage
%%%%%%%%%%%%%%%%%%%%%%%%%%%%%%%%%%%
 
\section*{Acknowledgments}

One of us (AW)  would like to thank Eric W. Hirschmann for useful
conversations and comments on gravitational collapse, and Nigel
Goldenfeld for his kind invitation and valuable discussions and interest
on critical phenomena in gravitational collapse. He would also like to express
his gratitude to the Department of Physics, UIUC, for hospitality.
The financial assistance from CAPES (AW) and CNPq (AYM) is gratefully 
acknowledged.

\end{document}